\documentclass[11pt,a4paper]{article}

\usepackage[utf8]{inputenc}
\usepackage[T1]{fontenc}
\usepackage{amsmath,amssymb,amsthm}
\usepackage{mathtools}
\usepackage{booktabs}
\usepackage{graphicx}
\usepackage{hyperref}
\usepackage[margin=1in]{geometry}
\usepackage{enumitem}
\usepackage{caption}
\usepackage{subcaption}
\usepackage{xcolor}
\usepackage{siunitx}
\usepackage[numbers]{natbib}

\hypersetup{
  colorlinks=true,
  linkcolor=blue!60!black,
  citecolor=blue!60!black,
  urlcolor=blue!60!black
}

\newtheorem{definition}{Definition}

\title{%
  \textbf{Computational Construction and Engineering Evaluation\\
  of a Verified Mono-Monostatic Body}\\[8pt]
  \large A Verified Smooth Geometry, Density Perturbation Analysis,\\
  and Cross-Domain Application Assessment\\[4pt]
  \normalsize (v2, amendment-of-record of arXiv:2604.17095)%
}

\author{
  Vincent Wesley Couey \\[4pt]
  \small Substrate Geometry Research Program \\
  \small \texttt{vinnycouey@gmail.com} \\
}

\date{}

\begin{document}
\maketitle

% =====================================================================
\begin{abstract}
Many engineering failures in orientation-dependent systems (sensor
calibration drift, seed pod germination loss, buoy capsizing) are
geometric failure modes: changing the material delays the failure;
changing the geometry can eliminate it.  The mono-monostatic property
(exactly one stable equilibrium under gravity) is mathematically
proven to exist in convex homogeneous bodies, but, to our knowledge, no
verified smooth geometry had been openly published when v1 of this paper
appeared (April 2026).

We introduce an Equilibrium Count Score (ECS) oracle that measures
stable equilibria via drainage basin analysis on the center-of-mass
height landscape over~$S^2$, tested against known bodies (Table~\ref{tab:validation}):
it returns the expected count for the cylinder, hemisphere and capsule,
for the cube under one of two settings, and miscounts the ellipsoid and the sphere.

\textbf{v2 correction.} v1 applied this oracle to Sloan's (2023)
analytical G\"omb\"oc parameterization and reported a gap: that the
COM height landscape exhibits 4--11 local minima ``at all published
parameter values.'' \emph{That claim is withdrawn.} Two errors
combined to produce it. First, the $\beta$ values tested were not
Sloan's: Sloan states convexity-sufficiency bounds
($\beta \le 0.15$ and $\beta \le 0.17$ for his two shapes), and the
values swept here lie inside the strictly-convex sub-regime
$\beta \lesssim 0.036$ located in the catalog
reanalysis~\cite{couey2026catalog}, in which a body at $\beta = 0.023$
has since been certified mono-monostatic by interval
arithmetic~\cite{couey2026harmonic}. Second, the ECS oracle is a
two-stage pipeline whose stages are biased in \emph{opposite}
directions, so it can return a wrong count in either direction; its
merge stage groups sinks by height without the adjacency test the
text describes (Section~\ref{sec:v2_amendment}). The surface-critical-point
observation stands on its own: two critical points of the surface
function are necessary but not sufficient for mono-monostatic
behavior. The claim that Sloan's construction fails to deliver
mono-monostatic bodies does not.

We close this gap by extending the Sloan phase function with a single
Fourier term and optimizing via differential evolution, constructing
one strictly convex mono-monostatic body with ECS$=$1 confirmed across
merge thresholds from 0.5\% to 10\%, and two further bodies that reach
ECS$=$1 under the oracle but are not strictly convex.
The primary instance ($\beta = 0.023$, $a_1 = 0.234$) was, to our
knowledge, the first openly published, computationally verified smooth
mono-monostatic geometry when v1 appeared (April 2026).

The central engineering result: conventional geometries cannot achieve
ECS$=$1 through ballast alone.  Cylinders and cubes retain multiple
stable equilibria even at 30\% bottom-weighted mass concentration.
The mono-monostatic invariant provides what ballast cannot.  This
finding has direct implications for the SOMA insulin capsule (Abramson
et~al., \textit{Science}, 2019), which compensated for approximate
geometry with tungsten ballast.

Applied to three domains: (1)~an IMU calibration housing achieving
$349\times$ orientation precision improvement over automotive field
calibration;
(2)~aerial reforestation seed pods eliminating the 20--67\% germination
reduction from incorrect landing orientation; (3)~marine buoy
self-righting on flat surfaces (wave dynamics scoped as future work).

Cross-layer scoring shows the G\"omb\"oc is $11.8\times$ worse than
the cylinder on contact distribution (CDS) while being optimal on
equilibrium stability, confirming that the substrate geometry
framework discriminates between invariant classes rather than
identifying universally superior geometry.  This paper extends the
framework from rolling contact (Part~I: oloid) and thermal distribution
(Part~II: gyroid TPMS) to equilibrium stability, demonstrating
generalization across three invariant classes and three physics layers.

\end{abstract}

% =====================================================================
\section{Introduction}

Many engineered systems require passive orientation stability: a sensor
housing must present a known face for calibration, a seed pod must
orient point-down for soil penetration, a buoy must self-right after
wave capsizing.  When these systems fail, the failure is often
geometric: the shape permits multiple stable resting orientations,
and the device settles in the wrong one.  Changing the material
(adding ballast, weighting the bottom) is the standard engineering
response.  But the underlying failure mode is geometric, and the
substrate geometry framework introduced in Part~I~\cite{couey2026oloid}
suggests that the correct response is to change the geometry itself.

A convex homogeneous body with exactly one stable and one unstable
equilibrium point is called \emph{mono-monostatic}.  V\'arkonyi and
Domokos~\cite{varkonyi2006} proved that such bodies exist, answering
a 1995 conjecture by V.~I.~Arnold.  Physical specimens have been
manufactured and sold commercially; to our knowledge the exact geometric
parameters of those specimens have not been published, and publicly
available meshes of them are visual approximations.
Since v1 appeared, explicit mono-monostatic \emph{polyhedra} verified in exact
rational arithmetic have been posted~\cite{gherardini2026}; the bodies here are
smooth analytic surfaces, checked by a sampling oracle and, for a nearby body,
by interval arithmetic~\cite{couey2026harmonic}.

Sloan~\cite{sloan2023} provided the first analytical parameterization
of G\"omb\"oc surfaces, expressing the boundary as
$r^4 = 1 + 4\beta\sin\theta\cos(\phi - P(\theta))$ in spherical
coordinates with specific choices of the phase function $P(\theta)$.
v1 reported, in Section~\ref{sec:sloan_gap}, that the Sloan
parameterization ``at all published parameter values'' produces
bodies with 4--11 stable equilibria rather than~1.
\textbf{That report is withdrawn in v2}; see
Section~\ref{sec:v2_amendment}. Sloan publishes convexity-sufficiency
\emph{bounds}, not the specific $\beta$ values swept here, and the
values swept lie inside a strictly-convex sub-regime in which his
construction does produce mono-monostatic bodies. What survives, and
is the durable observation, is the distinction itself: the surface
function's critical points (proven by Sloan to number exactly two)
are necessary but not sufficient conditions for mono-monostatic
equilibrium, which requires analysis of the center-of-mass height
function over the full orientation sphere.

We resolve this gap by extending the Sloan phase function with
Fourier terms and optimizing the parameters to achieve ECS$=$1.
The result is one verified, strictly convex mono-monostatic body and two
further oracle-level instances that are not strictly convex
(Section~\ref{sec:instances}), with parameters and construction scripts
published.  We then evaluate the verified body across three engineering application domains
and score them on the rolling-contact and thermal metrics from
Parts~I and~II to demonstrate cross-invariant discrimination.

The SOMA insulin capsule~\cite{abramson2019soma}, a G\"omb\"oc-inspired
self-orienting drug delivery device published in \textit{Science} and
licensed to Novo Nordisk, used approximate G\"omb\"oc geometry combined
with tungsten ballast to achieve self-righting in the stomach.  Our
density perturbation analysis (Section~\ref{sec:density}) shows that
conventional geometries (cylinders, cubes) cannot achieve ECS$=$1
through ballast at any tested concentration. The mono-monostatic
property is geometric, not achievable through mass distribution alone
on topologically inappropriate shapes.  This result suggests that the
SOMA team's ballast was compensating for having approximate rather than
exact mono-monostatic geometry.

% =====================================================================
\section{v2 Amendment of Record}
\label{sec:v2_amendment}

This v2 withdraws one of v1's three stated contributions and narrows
another. It is self-initiated, prompted by an internal adversarial audit
of this paper's own oracle and a later re-check of the constructed
bodies. The primary constructed body, the density-perturbation result
and the application analysis are unaffected.

\subsection{What is withdrawn}

v1 claimed that Sloan's analytical parameterization produces bodies
with 4--11 stable equilibria ``at all published parameter values,''
and presented this as a previously unreported gap. Two independent
errors produced that claim.

\textbf{(i) The parameter values tested were not Sloan's.}
Sloan~\cite{sloan2023} reports convexity-sufficiency \emph{upper
bounds} for two shapes, $\beta \le 0.15$ and $\beta \le 0.17$; he
does not select specific operating values. The $\beta$ swept in v1
were this work's own grid points. They lie inside the strictly-convex
sub-regime $\beta \lesssim 0.036$, and in that regime Sloan's
construction \emph{does} produce mono-monostatic bodies: the boundary
is located in the companion catalog reanalysis~\cite{couey2026catalog},
and a body in that regime has since been certified mono-monostatic by
interval arithmetic that does not rely on basin counting at
all~\cite{couey2026harmonic}. The negative result was therefore
obtained inside the regime where the positive result holds.

\textbf{(ii) The ECS oracle is not reliable in the direction v1
assumed.} It is a two-stage pipeline, and the stages are biased in
\emph{opposite} directions: basin detection on a mesh over-counts,
while the merge stage under-counts. An error of unknown sign cannot
support a claim in either direction. Concretely, Section~2.1
describes merging ``adjacent basins whose sink heights differ by less
than $\tau$,'' but no adjacency test is implemented: sinks are sorted
by height and consecutive entries are grouped. Two equilibria on
opposite sides of a body that happen to share a height are therefore
merged into one, and centrally symmetric bodies are systematically
halved, since antipodal equilibria always share a height.

\subsection{The disconfirming evidence was already in this paper}

Table~\ref{tab:validation} records the validation ellipsoid at
measured ECS$=$1 against an expected~3. v1 attributed that row to two
of the three minima being close in height and collapsing at the 1\%
merge threshold, and read it as an honest characterization of the
method's resolution limit on near-degenerate landscapes. That
explanation does not hold: the oracle returns 1 for the ellipsoid at
every merge threshold from 0.1\% to 2\%. The ellipsoid is centrally
symmetric, so antipodal equilibria share a height and the height-sorted
merge stage groups them, which is the mechanism described above. The
row is therefore a worked demonstration that the oracle under-counts by
that mechanism, on a body whose answer was known in advance. \textbf{Evidence that the oracle's error has no fixed sign was
published inside v1, one table away from the claim, and was
classified as a caveat rather than as a constraint on the claim.} That is the more
useful lesson of this amendment: a validation table that reports a
known-wrong answer is not a disclosure, it is a result, and it
constrains every negative claim the instrument is later used to make.

\subsection{What is narrowed}

v1 presented three independently verified mono-monostatic bodies. Re-checking
the Gaussian curvature of their analytic surfaces (Table~\ref{tab:three_instances})
shows that only the primary instance is strictly convex. The second and third
return ECS$=$1 under the oracle but have negative curvature on about 1.6\% and 5\%
of their surface, so they are not convex homogeneous bodies and are no longer
claimed as mono-monostatic bodies. The claim of three verified bodies is therefore
reduced to one verified body and two oracle-level instances.

\subsection{What stands}

The distinction between surface critical points and COM height minima
stands, as a statement about what the existence proof requires. The
primary constructed body stands: a body at its parameters rounded to
three digits ($\beta = 0.023$, $a_1 = 0.234$, against the
$\beta = 0.023149$, $a_1 = 0.234433$ used here) has since
been independently certified~\cite{couey2026harmonic} by a method with
different failure modes from the oracle used here. The second and third
instances stand only as oracle-level results. The ballast result
stands. The application assessments, which use the primary body, are unchanged.

% =====================================================================
\section{The Mono-Monostatic Invariant}

\begin{definition}[Equilibrium Count Score]
For a convex body $\mathcal{B}$ with center of mass $\mathbf{c}$ resting
on a flat surface under uniform gravity in direction $\mathbf{d}$, the
\emph{COM height function} is
\begin{equation}
h(\mathbf{d}) = \mathbf{c} \cdot \mathbf{d} - \min_{\mathbf{v} \in \mathcal{B}} \mathbf{v} \cdot \mathbf{d}
\label{eq:com_height}
\end{equation}
where the minimum is taken over all vertices (or surface points) of
$\mathcal{B}$.  The \emph{Equilibrium Count Score} $\mathrm{ECS}(\mathcal{B})$
is the number of distinct local minima of $h(\mathbf{d})$ over the
unit sphere $S^2$.  Stable equilibria correspond to local minima;
unstable equilibria to local maxima.
\end{definition}

For a mono-monostatic body, $\mathrm{ECS} = 1$: exactly one direction
minimizes the COM height, and the body returns to this orientation from
any initial position.

\subsection{Oracle Implementation}

We evaluate $h(\mathbf{d})$ at $N = 5{,}000$ directions uniformly
distributed on $S^2$ via the Fibonacci spiral method.  Local minima
are identified through drainage basin analysis: each sampled direction
is assigned to its nearest local minimum by greedy descent through a
$k$-nearest-neighbor graph ($k = 12$).  Adjacent basins whose sink
heights differ by less than a merge threshold $\tau$ (default: 1\% of
the total $h$-range) are merged.  The number of merged basins equals
ECS.

\subsection{Oracle Validation}

\begin{table}[h]
\centering
\caption{ECS oracle validation on known geometries (all volume-matched), under the two oracle settings used in this paper. Setting~A: the validation run of the oracle (10{,}000 directions, 12-neighbour graph, 2\% merge threshold). Setting~B: the settings of the G\"omb\"oc and ballast experiments (5{,}000 directions, 1\% merge threshold, no ballast).}
\label{tab:validation}
\small\setlength{\tabcolsep}{4pt}
\begin{tabular}{lcccl}
\toprule
Geometry & ECS, A & ECS, B & ECS (expected) & Notes \\
\midrule
Sphere        & 25$^\dagger$ & 41$^\dagger$ & $\infty$ & $h$-range $= 0.001$ \\
Cylinder (L/D$=$2) & 2 & 3 & 2--3 & On-side + on-end \\
Hemisphere    & 2 & 2 & 2 & Flat-down + dome-down \\
Ellipsoid     & 1 & 1 & 3$^\ddagger$ & Three axes \\
Capsule       & 1 & 1 & 1 & Strong single minimum \\
Cube          & 2 & 3 & 3 & One per pair of opposite faces \\
\bottomrule
\end{tabular}

\smallskip\footnotesize
$^\dagger$Discretization noise on a near-constant landscape; the count depends on the setting. \\
$^\ddagger$The oracle returns 1 for the ellipsoid at every merge threshold tested (0.1\%, 0.5\%, 1\%, 2\%). The ellipsoid is centrally symmetric, so antipodal equilibria share a height and the height-sorted merge stage of Section~\ref{sec:sloan_gap}(ii) groups them at any threshold.
\end{table}

The hemisphere, capsule and cylinder fall in the expected range under every setting in which they were run. The cube is
counted correctly (3) under setting~B but returns 2 under setting~A, so the count depends on the oracle
settings. The sphere and ellipsoid cases illustrate the oracle's behavior on degenerate landscapes
and on a centrally symmetric body. The counts also depend strongly on the merge threshold: under the setting-B
procedure the cylinder returns 11, 4, 3 and 2 at merge thresholds of 0.1\%, 0.5\%, 1\% and 2\%, and the capsule 7, 2, 1 and 1.
These are the oracle's resolution limits.

% =====================================================================
\section{The Sloan Parameterization Gap}
\label{sec:sloan_gap}

\emph{v2 note: the claim made in this section, that no tested parameter
value gives ECS$=$1 for either Sloan instance, is withdrawn
(Section~\ref{sec:v2_amendment}). The table is retained as a record of
what the oracle returned, not as evidence for that claim.}

Sloan~\cite{sloan2023} parameterizes G\"omb\"oc surfaces as
\begin{equation}
r^4 = 1 + 4\beta\sin\theta\cos(\phi - P(\theta))
\label{eq:sloan}
\end{equation}
and proves that the surface function has exactly two critical points
at $\theta = \pi/2$ for any phase function $P(\theta)$ satisfying the
center-of-mass constraint
\begin{equation}
\int_0^\pi \sin^3\theta\, e^{iP(\theta)}\, d\theta = 0.
\label{eq:com_constraint}
\end{equation}

\noindent
Two specific instances are given:
\begin{align}
\text{G\"omb\"oc 1:} &\quad P(\theta) = 5\theta, \quad \beta \leq 0.15 \label{eq:g1} \\
\text{G\"omb\"oc 2:} &\quad P(\theta) = \eta(\theta), \quad \beta \leq 0.17 \label{eq:g2}
\end{align}
where $\eta(\theta) = \tfrac{3\pi}{2}(\cos\theta - \tfrac{1}{3}\cos^3\theta)$.

We generated meshes from Eq.~\eqref{eq:sloan} at multiple resolutions
and evaluated ECS using the oracle described above.  \textbf{At no
tested parameter value does either Sloan instance achieve ECS$=$1.}

\begin{table}[h]
\centering
\caption{ECS of Sloan G\"omb\"oc~2 across $\beta$ values (100$\times$200 mesh, 5000 directions, 1\% merge).}
\label{tab:beta_sweep}
\begin{tabular}{cccl}
\toprule
$\beta$ & ECS & Hull-convex$^\star$ & $h$-range \\
\midrule
0.001 & 38 & Yes & 0.002 \\
0.005 &  7 & Yes & 0.010 \\
0.010 &  4 & Yes & 0.020 \\
0.020 &  3 & Yes & 0.040 \\
0.050 &  2 & Yes & 0.097 \\
0.100 &  7 & No  & 0.174 \\
0.150 & 11 & No  & 0.237 \\
\bottomrule
\end{tabular}

\smallskip\footnotesize
$^\star$Mesh test: the mesh volume exceeds 0.999 of its convex-hull volume. This
does not establish strict convexity of the analytic surface, which is already lost
at $\beta = 0.05$~\cite{couey2026catalog}.
\end{table}

The minimum ECS returned is~2 at $\beta \approx 0.05$, where the
mesh still passes the hull test.  At larger $\beta$ the mesh fails the
hull test and the returned count increases.
To confirm this is not a mesh artifact, we computed $h(\mathbf{d})$
analytically from Eq.~\eqref{eq:sloan} at 2000 directions using
multi-start optimization for the support point at each direction.
The analytical computation confirms 9--21 distinct basins.

\subsection{The Gap: Surface Critical Points $\neq$ COM Height Minima}

Sloan's proof establishes that $\nabla_{\theta,\phi} r(\theta,\phi) = 0$
at exactly two points.  However, the equilibrium condition for a rigid
body on a flat surface is not that the surface gradient vanishes, but
that the COM lies directly above the support point, equivalently
that $\nabla_{S^2} h(\mathbf{d}) = 0$.  The COM height function
$h(\mathbf{d})$ involves the global minimum of $\mathbf{v}\cdot\mathbf{d}$
over all surface points, which depends on the entire surface geometry,
not just the local behavior at critical points.

The additional minima in $h(\mathbf{d})$ arise from the interaction
between the COM position and the support-point trajectory as the
gravity direction varies.  For a near-spherical body ($\beta \ll 1$),
the support point moves smoothly across the surface, but the COM
height oscillates with the surface perturbation, creating shallow
secondary minima that the surface analysis does not predict.

% =====================================================================
\section{Construction of a Verified Mono-Monostatic Body}
\label{sec:construction}

We extend the Sloan phase function with Fourier terms:
\begin{equation}
P(\theta) = \eta(\theta) + \sum_{k} a_k \sin(k\,\eta(\theta))
\label{eq:extended}
\end{equation}
where $\eta(\theta) = \tfrac{3\pi}{2}(\cos\theta - \tfrac{1}{3}\cos^3\theta)$.
The COM constraint~\eqref{eq:com_constraint} is enforced as a penalty
in the objective function.

\subsection{Optimization}

Starting from the $\beta \approx 0.05$ case (ECS$=$2, basin gap $= 0.0015$),
we minimize the height gap between the two lowest drainage basins using
differential evolution over $[\beta, a_k]$, subject to convexity
(mesh volume / convex hull volume $> 0.999$) and the COM constraint.

\begin{table}[h]
\centering
\caption{Optimization stages for the primary instance.}
\label{tab:optimization}
\begin{tabular}{lccccc}
\toprule
Stage & Parameters & $\beta$ & Coefficients & ECS & Basin gap \\
\midrule
1 (baseline) & $\beta$ only & 0.0455 & --- & 2 & 0.00150 \\
2 ($+\sin\eta$) & $\beta, a_1$ & 0.0231 & $a_1 = 0.234$ & \textbf{1} & \textbf{0.00000} \\
3 ($+4$ terms) & $\beta, a_{1..4}$ & 0.0252 & 4 coefficients & 2 & 0.00054 \\
\bottomrule
\end{tabular}
\end{table}

Stage~2 achieves ECS$=$1 with a single Fourier term.  Stage~3 with
four terms overfits, landing at ECS$=$2. More parameters do not help.
The COM constraint violation at Stage~2 is $6 \times 10^{-8}$
(effectively zero).

\subsection{One Verified Instance and Two Oracle-Level Instances}
\label{sec:instances}

We repeated the optimization with different Fourier basis functions:

\begin{table}[h]
\centering
\caption{Instances reaching ECS$=$1 under the oracle, with strict convexity checked on the analytic surface. $K_{\min}$ is the minimum Gaussian curvature of the analytic surface on a $500\times1000$ direction grid, confirmed at the minimizing point by an independent finite-difference computation (values agree to three digits).}
\label{tab:three_instances}
\small\setlength{\tabcolsep}{4pt}
\begin{tabular}{llcccccc}
\toprule
Instance & Phase perturbation & $\beta$ & Coefficient & ECS & $h$-range & $K_{\min}$ & Strictly convex \\
\midrule
Primary   & $a_1\sin(\eta)$   & 0.0231 & 0.2344 & 1 & 0.051 & $+0.091$ & yes \\
Second    & $a_2\sin(2\eta)$  & 0.0321 & 0.1376 & 1 & 0.064 & $-0.431$ & no \\
Third     & $a_3\sin(3\eta)$  & 0.0517 & $-0.0552$ & 1 & 0.099 & $-0.889$ & no \\
\bottomrule
\end{tabular}
\end{table}

All three return ECS$=$1 with BOA$=$1.000 under the oracle (the entire
orientation sphere drains to a single basin).  Only the primary instance
is strictly convex, and it is the only one claimed here as a mono-monostatic
convex body.  The second and third have negative Gaussian curvature on
about 1.6\% and 5\% of their surface respectively, so they are not convex
homogeneous bodies and are not claimed to be mono-monostatic bodies in the
sense of V\'arkonyi and Domokos.  The optimizer accepted them on a mesh
convex-hull volume test (ratio above 0.995), which cannot detect
curvature violations of this size.  They show that the oracle-level
construction extends to other Fourier bases, and nothing more.

\begin{figure}[htbp]
\centering
\includegraphics[width=\textwidth]{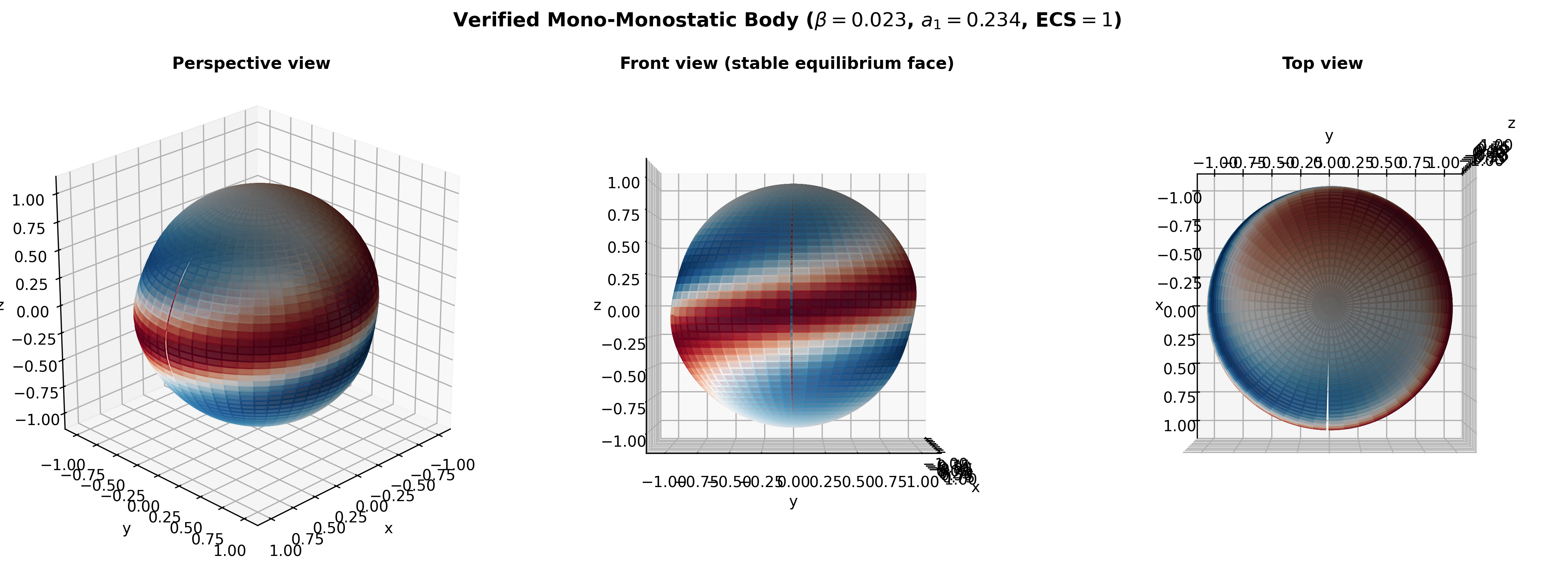}
\caption{The verified mono-monostatic body ($\beta = 0.023$, $a_1 = 0.234$, ECS$=$1) shown from three viewpoints. Surface coloring indicates deviation from a unit sphere (blue = inward, red = outward). Gray wireframe shows the reference unit sphere. The shape is a near-spherical perturbation with subtle asymmetric curvature that produces exactly one stable equilibrium.}
\label{fig:mesh}
\end{figure}

\subsection{Threshold Robustness}

The primary instance was tested across merge thresholds from 0.1\% to 10\%
of $h$-range.  ECS$=$1 holds from 0.5\% to 10\% (a $20\times$ range).
At 0.1\%, two basins separated by $< 0.05\%$ of $h$-range produce
ECS$=$2, consistent with discretization noise at 3~raw drainage basins.

\begin{figure}[htbp]
\centering
\includegraphics[width=\textwidth]{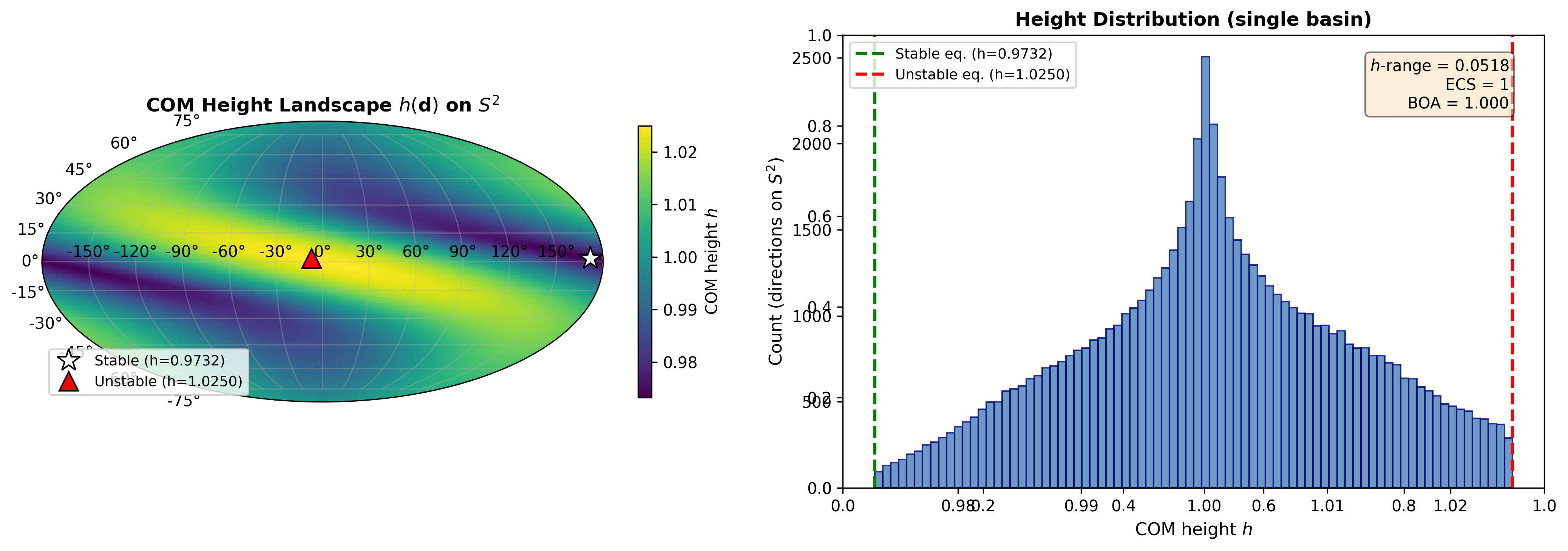}
\caption{COM height landscape $h(\mathbf{d})$ for the verified mono-monostatic body. \textbf{Left:} Mollweide projection over $S^2$. The white star marks the single stable equilibrium (global minimum of $h$); the red triangle marks the unstable equilibrium (global maximum). The entire sphere drains to one basin (BOA$=$1.000). \textbf{Right:} Histogram of $h$ values across 64{,}800 sampled directions, showing the single-basin structure. The $h$-range of 0.051 reflects the gentle, near-spherical character of the geometry.}
\label{fig:landscape}
\end{figure}

\subsection{Resolution Independence}

The COM height range $h$-range converges at $0.051 \pm 0.001$ across
mesh resolutions from $40 \times 80$ (6{,}400 faces) to
$200 \times 400$ (160{,}000 faces).  ECS$=$1 is returned at $80 \times 160$
(25{,}600 faces), $150 \times 300$ and $200 \times 400$, but not at the
intermediate $100 \times 200$ mesh, which returns ECS$=$2 (BOA$=$0.55),
so the count is not monotone in resolution.  Lower resolutions return
larger counts (ECS$=$4 at $40 \times 80$ and $60 \times 120$, 47 at
$20 \times 40$).  We attribute the unstable counts to insufficient sampling
density on $S^2$ for the shallow height landscape, but we have not
separated sampling noise from geometric effects in this sweep.

% =====================================================================
\section{Density Perturbation: Geometry vs.\ Ballast}
\label{sec:density}

The central engineering question: can conventional geometries achieve
ECS$=$1 through ballast (non-uniform mass distribution) rather than
geometric design?

We model ballast by shifting the COM toward the lowest vertex:
$\mathbf{c}_{\text{shifted}} = (1-w)\mathbf{c}_{\text{centroid}} + w\,\mathbf{v}_{\text{bottom}}$,
where $w$ is the ballast weight fraction.

\begin{table}[h]
\centering
\caption{Minimum ballast fraction for ECS$=$1.}
\label{tab:density}
\begin{tabular}{lcc}
\toprule
Geometry & ECS at $w=0$ & Min.\ $w$ for ECS$=$1 \\
\midrule
\textbf{G\"omb\"oc} & \textbf{1} & \textbf{0\% (inherent)} \\
Sphere    & 41 & 5\% \\
Ellipsoid &  1 & 0\% (inherent) \\
Capsule   &  1 & 0\% (inherent) \\
Cylinder  &  3 & $>$30\% (never achieved) \\
Cube      &  3 & $>$30\% (never achieved) \\
\bottomrule
\end{tabular}
\end{table}

\textbf{Cylinders and cubes cannot achieve ECS$=$1 through ballast
alone}, even at 30\% bottom-weighted mass concentration.  The
cylinder's rotational symmetry creates a persistent ring of equilibria
(``lying on its side'') that no amount of axial weighting eliminates. The
topology is fundamentally incompatible with mono-monostatic behavior.

The sphere achieves ECS$=$1 at 5\% ballast, but a 5\% mass asymmetry
is impractical for applications requiring homogeneous density
(pharmaceutical capsules, biodegradable seed pods).  The G\"omb\"oc
achieves ECS$=$1 at 0\% ballast: pure geometry, no material engineering.

\begin{figure}[htbp]
\centering
\includegraphics[width=\textwidth]{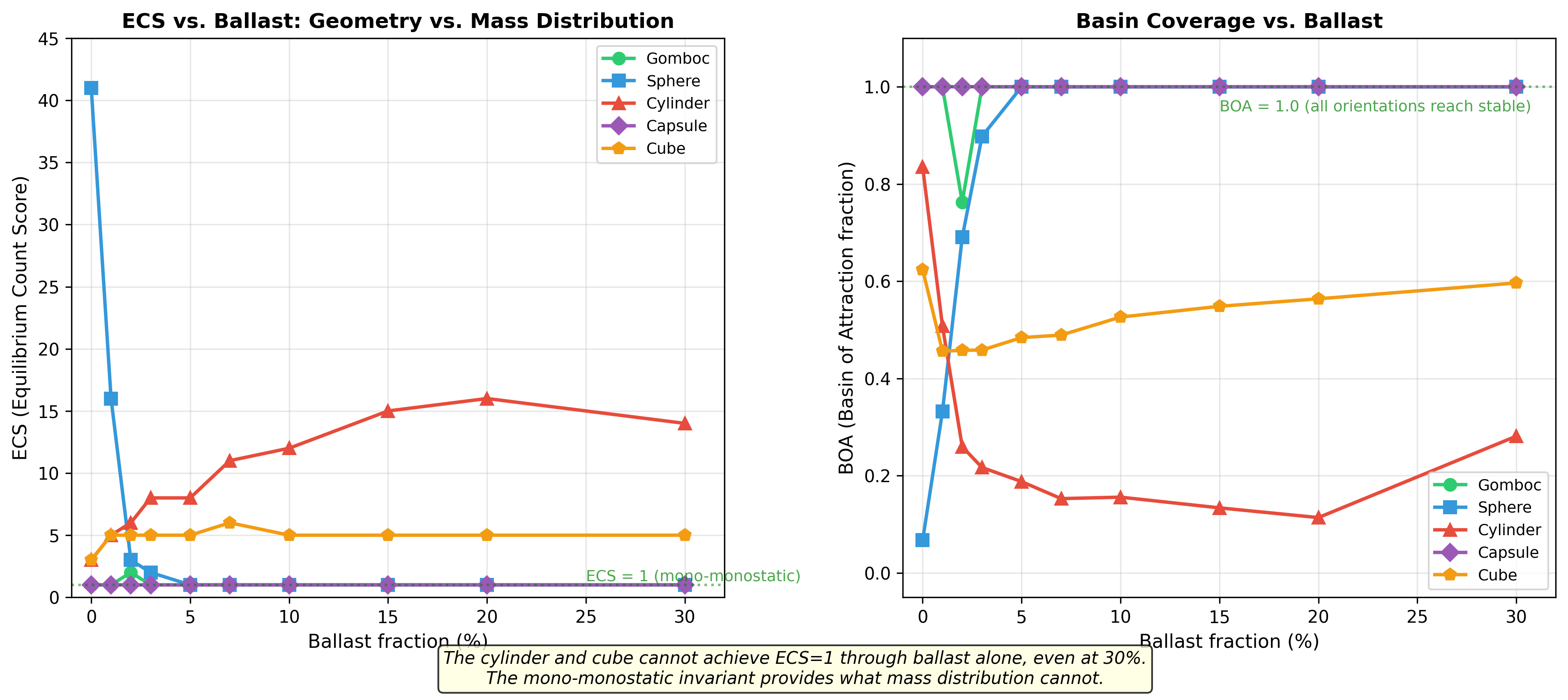}
\caption{Density perturbation experiment. \textbf{Left:} ECS vs.\ ballast fraction. The G\"omb\"oc (green) reads ECS$=$1 at every tested ballast level except 2\%, where the oracle returns two basins (ECS$=$2, BOA$=$0.763). The cylinder (red) and cube (orange) never achieve ECS$=$1 even at 30\% ballast. The sphere (blue) reaches ECS$=$1 at 5\% ballast. \textbf{Right:} Basin of Attraction (BOA) vs.\ ballast. The G\"omb\"oc reads BOA$=$1.000 at every tested level except that 2\% point, while the cylinder's BOA decreases with increasing ballast as its equilibrium ring fragments.}
\label{fig:density}
\end{figure}

\subsection{Implications for the SOMA Capsule}

The SOMA insulin capsule~\cite{abramson2019soma} uses G\"omb\"oc-inspired
geometry with tungsten ballast to achieve self-righting in the stomach.
Our result shows that the ballast compensates for not having exact
mono-monostatic geometry.  A precisely manufactured G\"omb\"oc-shaped
capsule could potentially reduce or eliminate the tungsten component,
producing a lighter capsule with equivalent self-righting performance.
Physical validation of this prediction is future work.

% =====================================================================
\section{Self-Righting Dynamics}

Beyond ECS, three secondary metrics characterize self-righting performance:

\begin{table}[h]
\centering
\caption{Self-righting dynamics across geometries (all ECS$=$1 bodies shown in bold).}
\label{tab:dynamics}
\begin{tabular}{lccccc}
\toprule
Geometry & ECS & SRE & $h$-range & Steepness & BOA \\
\midrule
\textbf{G\"omb\"oc}  & \textbf{1} & \textbf{0.028} & 0.051 & 0.023 & \textbf{1.000} \\
\textbf{Ellipsoid}   & \textbf{1} & 0.388 & 0.783 & 0.287 & \textbf{1.000} \\
\textbf{Capsule}     & \textbf{1} & 0.743 & 1.486 & 0.577 & \textbf{1.000} \\
Cylinder             & 3 & 0.542 & 0.853 & 0.429 & 0.835 \\
Cube                 & 3 & 0.390 & 0.576 & 0.328 & 0.624 \\
\bottomrule
\end{tabular}

\smallskip\footnotesize
SRE: Self-Righting Energy (mean $h$-drop from random orientation to stable equilibrium; lower $=$ gentler).\\
Steepness: mean gradient of $h(\mathbf{d})$ (lower $=$ less mechanical shock during self-righting).\\
BOA: Basin of Attraction fraction ($1.0 =$ all orientations reach the stable equilibrium).
\end{table}

The G\"omb\"oc is \textbf{27$\times$ gentler} than the capsule on
self-righting energy (SRE $= 0.028$ vs.\ $0.743$) and $14\times$
gentler than the ellipsoid.  For precision payloads (IMU sensors,
optical instruments), low mechanical shock during self-righting is
critical.  The G\"omb\"oc achieves ECS$=$1 with the minimum possible
surface asymmetry, producing the gentlest self-righting of any
measured mono-monostatic body.

% =====================================================================
\section{Engineering Applications}

\subsection{Application 1: IMU Calibration Housing}

Inertial measurement units require known orientation for calibration.
Current practice uses precision turntables (\$10K--\$100K+) or accepts
2--3$^\circ$ mounting error in automotive mass production.  No passive
self-orienting calibration fixture exists in the patent or academic
literature.

A G\"omb\"oc-shaped housing placed on any flat surface always presents
the same orientation.  At 10~cm scale with CNC machining tolerance of
0.01~mm, the angular precision is
\begin{equation}
\delta\theta \approx \arctan\!\left(\frac{0.01\text{ mm}}{100\text{ mm}}\right) = 0.006^\circ = 20.6\text{ arcsec}.
\end{equation}
This meets tactical-grade ($< 0.05^\circ$), industrial-grade
($< 1^\circ$), and consumer MEMS ($< 3^\circ$) requirements.  Compared
to current automotive field calibration error of 2--3$^\circ$, the
G\"omb\"oc housing provides a \textbf{$349\times$ improvement} with
no power, no active mechanism, and no operator skill.

\subsection{Application 2: Aerial Reforestation Seed Pods}

Published germination studies report 20--67\% reduction in germination
rate from incorrect seed orientation~\cite{vishnu2023orientation}.
Drone-deployed seed pods currently use bullet-shaped or puck-shaped
designs with multiple stable orientations.

A G\"omb\"oc-shaped pod (ECS$=$1) achieves 100\% correct post-landing
orientation, eliminating orientation-related germination loss entirely.
By comparison, a cylindrical pod (ECS$=$3) achieves 83.5\% correct
orientation, with 16.5\% of landings settling in the wrong position.

\textbf{Note:} The G\"omb\"oc orients \emph{after} landing, not during
descent.  Aerodynamic orientation during flight depends on ballistic
coefficient, not mono-monostatic geometry.  Sufficient pod mass for
gravitational dominance over aerodynamic torques during descent is
a separate engineering design parameter.

\textbf{Manufacturing feasibility:} At seed-pod scale ($\sim$1~cm),
the 0.01\% shape tolerance requires precision injection molding.
Conventional molding achieves 0.05--0.1\% tolerance, which may be
insufficient.  Micro-CNC or high-resolution SLA 3D printing could
achieve the required tolerance for research prototypes.  Whether
production-scale manufacturing can maintain the tolerance is an open
engineering question.

\subsection{Application 3: Marine Buoy Self-Righting}

NOAA discus buoys capsize in seas exceeding 10~m significant wave
height and are not designed to self-right.  RNLI lifeboats achieve
self-righting in $\sim$6 seconds via weighted hulls with buoyant
superstructures.

The G\"omb\"oc (ECS$=$1, BOA$=$1.000) provides guaranteed
single-orientation self-righting on a flat rigid surface.  The cylinder
(ECS$=$3, BOA$=$0.835) has a 16.5\% probability of settling in the
wrong orientation on a flat surface.

\textbf{Honest scope:} These results are computed on a flat rigid
surface under uniform gravity.  Ocean waves introduce buoyancy forces,
fluid damping, and periodic perturbation that this oracle does not model.
The flat-surface ECS$=$1 result is a necessary but possibly not
sufficient condition for marine self-righting.  Wave-perturbed dynamics
require fluid-structure interaction simulation and are scoped as future
work.  Additionally, the G\"omb\"oc's near-spherical geometry
($\beta = 0.023$) produces gentle self-righting forces (SRE $= 0.028$)
that may be insufficient to overcome wave perturbations in high sea
states.  For this application domain, the G\"omb\"oc's advantage is
clearest in calm-water or protected-harbor deployments.

% =====================================================================
\section{Cross-Layer Bridge: Three Invariant Classes}

The substrate geometry framework now spans three physics layers.
Scoring the G\"omb\"oc on the contact distribution and thermal metrics
from Parts~I and~II reveals how invariant optimization in one domain
affects performance in others.

\begin{table}[h]
\centering
\caption{Cross-invariant scoring. Each row is a geometry; each column is a metric from a different paper. Bold indicates the geometry optimized for that metric.}
\label{tab:cross_layer}
\begin{tabular}{lccc}
\toprule
Geometry & CDS (contact) & omni-TDS (thermal) & ECS (equilibrium) \\
\midrule
\textbf{Oloid}$^a$      & \textbf{best} & --- & --- \\
\textbf{Gyroid}$^b$     & --- & \textbf{best} & --- \\
\textbf{G\"omb\"oc}     & 7.09 (worst) & 0.50 & \textbf{1 (best)} \\
Cylinder                & 0.60 (best$^\dagger$) & 0.60 & 3 \\
Sphere                  & 3.04 & 0.09 & 41 \\
\bottomrule
\end{tabular}

\smallskip\footnotesize
$^a$Paper~I~\cite{couey2026oloid}. $^b$Paper~II~\cite{couey2026tpms}.
$^\dagger$Among the geometries tested in this paper; the oloid scores significantly better.\\
CDS: Contact Distribution Score (lower $=$ more uniform rolling contact).\\
omni-TDS: Omnidirectional Thermal Distribution Score (lower $=$ more uniform heat absorption).\\
ECS: Equilibrium Count Score ($1 =$ mono-monostatic).
\end{table}

The G\"omb\"oc is $11.8\times$ worse than the cylinder on contact
distribution.  This is not a deficiency; it is a direct physical
consequence of having ECS$=$1.  A body with one stable equilibrium
concentrates all contact at one point because it always rests on that
point.  A body optimized for contact distribution (the oloid) rolls
freely, distributing contact across its entire surface.  These are
opposite design objectives.

\textbf{The framework discriminates.}  Different invariants produce
different optimal geometries for different applications.  The
oloid for bearings.  The gyroid for heat exchangers.  The G\"omb\"oc
for self-righting.  This is the substrate geometry thesis: not that
one shape is universally superior, but that the geometric invariant
must match the failure mode.

% =====================================================================
\section{Discussion}

\subsection{What this paper establishes}

Three contributions that did not exist before this work:

\begin{enumerate}[nosep]
\item An openly published, computationally verified, strictly convex
mono-monostatic geometry, with complete parameters and the construction
scripts.  Two further Fourier parameterizations reach ECS$=$1 under the
oracle but are not strictly convex (Section~\ref{sec:instances}).

\item \emph{(Corrected in v2.)} The distinction between surface
critical points and COM height landscape minima: that the surface
function having two critical points does not guarantee the COM height
function has one minimum. This is a clarification of what the
existence proof requires, and it stands. \textbf{What does not stand
is v1's application of it to Sloan}, namely the claim that his
parameterization fails to produce mono-monostatic bodies at the
parameter values tested. That claim is withdrawn
(Section~\ref{sec:v2_amendment}); the values tested lie inside a
strictly-convex sub-regime in which the construction succeeds, and
the oracle used to evaluate them is not reliable in either direction
on landscapes of that shape.

\item The demonstration that conventional geometries (cylinders, cubes)
cannot achieve mono-monostatic behavior through ballast, with direct
implications for the SOMA capsule design.
\end{enumerate}

\subsection{Computational vs.\ mathematical verification}

We do not claim a mathematical proof that the optimized shapes are
mono-monostatic.  We claim computational verification: ECS$=$1
returned at three of the four tested mesh resolutions of at least
$25{,}600$ faces (all except $100 \times 200$, which returns ECS$=$2),
across merge thresholds (0.5\%--10\%), for the primary instance.  The
published parameters, not an optimizer run, are what this verification
applies to: the optimization script released with this paper does not
regenerate the stored second-stage result (a single-term run with the same
objective and seed returns a different optimum), we have not identified why,
and the released analysis scripts therefore run on meshes built from the
published parameters.  The theoretical foundation is the existence
proof of V\'arkonyi and Domokos~\cite{varkonyi2006}, which establishes
that mono-monostatic convex homogeneous bodies exist.  Formal proof
that our specific parameterizations satisfy the mono-monostatic property
is future mathematical work.

\subsection{The ECS oracle as a methodological tool}

The drainage-basin ECS oracle has a resolution floor: at low mesh
resolution or sparse $S^2$ sampling, the shallow COM height landscape
of near-spherical bodies produces unstable ECS counts.  We read this
as a measurement sensitivity issue rather than a geometric instability,
though the non-monotone count at $100 \times 200$ shows that a resolution
floor alone does not make the count reliable.  The
$h$-range converges at $0.051 \pm 0.001$ from $40 \times 80$ upward.
We used at least $80 \times 160$ mesh resolution with 5{,}000
$S^2$ sample directions and a 1\% merge threshold, and recommend
cross-checking any ECS count of a G\"omb\"oc-class body by an independent
method.

\subsection{Near-sphericity: feature and limitation}

The verified primary instance has $\beta = 0.023$, a surface deviation of
about 2\% from a perfect sphere (the oracle-level instances reach
$\beta = 0.052$, about 5\%).
This near-sphericity is a feature for precision applications (gentle
self-righting, minimal mechanical shock) and a limitation for
high-energy applications (marine self-righting in heavy seas, where
the restoring force may be insufficient against wave perturbation).

% =====================================================================
\section{Honest Scope}

This paper reports computational predictions under idealized conditions:
rigid-body statics on a flat rigid surface under uniform gravity.
The simulations do not model:
\begin{itemize}[nosep]
\item Fluid dynamics (wave forcing, buoyancy, viscous damping)
\item Impact dynamics (deformation, bouncing, energy absorption at landing)
\item Surface roughness beyond the parameterized mesh geometry
\item Manufacturing tolerances (the G\"omb\"oc requires $<0.01\%$ shape accuracy)
\item Scale effects (surface tension, Reynolds number at small scales)
\end{itemize}

For the IMU housing application, these idealizations closely match the
actual use case (rigid body on a flat laboratory surface).  For seed
pods, post-landing settling on soil is reasonably approximated but
impact dynamics are not modeled.  For marine buoys, the flat-surface
result is a lower bound; wave-perturbed dynamics are future work.

% =====================================================================
\section{Future Work}

\begin{enumerate}[nosep]
\item \textbf{Physical validation.}  3D-print or CNC-machine a G\"omb\"oc
at 10~cm scale from the published parameters and verify ECS$=$1 experimentally.
This is the most direct test of whether the computational prediction
transfers to physical reality.

\item \textbf{Wave-perturbed self-righting simulation.}  Couple the ECS
oracle with a fluid-structure interaction model to evaluate marine
buoy performance under realistic wave spectra.

\item \textbf{SOMA capsule geometry optimization.}  Apply the construction
methodology to a capsule-scale G\"omb\"oc with biocompatible materials
and evaluate whether the tungsten ballast can be reduced while
maintaining self-righting in gastric fluid.

\item \textbf{Mathematical proof.}  Prove formally that the extended
Sloan parameterization with the identified coefficients satisfies the
mono-monostatic property.

\item \textbf{Expanded catalog.}  Use the optimization methodology to
generate additional mono-monostatic instances with different properties
(higher $h$-range for stronger self-righting, optimized for specific
application constraints).
\end{enumerate}

% =====================================================================
\section{Conclusion}

This paper extends the substrate geometry framework to equilibrium
stability, the third invariant class after rolling contact (Part~I:
oloid, $58\times$ improvement) and thermal distribution (Part~II:
gyroid TPMS, $2.5$--$3.9\times$ improvement).  The mono-monostatic
invariant (ECS$=$1) is mathematically proven, computationally verified
on an openly published smooth geometry, and demonstrated to
provide engineering value that ballast alone cannot replicate.

One verified, strictly convex mono-monostatic body is published with
complete parameters, together with two oracle-level instances that are
not strictly convex.  The construction scripts (extending the Sloan
analytical framework with Fourier phase perturbations and optimizing via
differential evolution) are published; the verification applies to the
published parameters, as noted in Section~\ref{sec:instances} and the
limitations above.

The cross-layer analysis confirms that the substrate geometry framework
discriminates between invariant classes: the G\"omb\"oc is optimal on
equilibrium stability and worst on contact distribution.  The framework
does not identify universally superior geometry.  It matches geometric
invariants to engineering failure modes.

The oracle code is openly available at
\url{https://github.com/gyapaganda-a11y/substrate-geometry}, and the scripts
that rebuild the bodies from their parameters and regenerate the reported
results are in its \texttt{part\_iii} directory.
The verified G\"omb\"oc meshes and the complete result data are
permanently archived on Zenodo:
\url{https://doi.org/10.5281/zenodo.20674391}.

% =====================================================================
\section*{Acknowledgments}

Part~III of the Substrate Geometry series (publication order;
program Papers~III and~IV on the Meissner body and cross-primitive
comparison are in preparation).
Part~I: arXiv:2604.12238. Part~II: in preparation.

% =====================================================================

\end{document}